\documentclass[11pt]{iopart}
\usepackage{iopams}  
\usepackage{graphicx}  	
\usepackage{color}     	
\usepackage{cite}		
\begin{document}

\title[Chemical potential in active systems]{Chemical potential in active systems: predicting phase equilibrium from bulk equations of state?}

\author{Siddharth Paliwal}
\ead{S.Paliwal@uu.nl}
\address{Soft Condensed Matter, Debye Institute for Nanomaterials Science, Utrecht University, Princetonplein 5, 3584 CC Utrecht, The Netherlands}
\author{Jeroen Rodenburg, Ren\'e van Roij}
\address{Institute for Theoretical Physics, Utrecht University, Princetonplein 5, 3584 CC Utrecht, The Netherlands}
\author{Marjolein Dijkstra}
\ead{M.Dijkstra@uu.nl}
\address{Soft Condensed Matter, Debye Institute for Nanomaterials Science, Utrecht University, Princetonplein 5, 3584 CC Utrecht, The Netherlands}

\date{\today}

\begin{abstract}
We derive a microscopic expression for a quantity $\mu$ that plays the role of chemical potential of Active Brownian Particles (ABPs) in a steady state in the absence of vortices.   We show that  $\mu$ consists of (i) an intrinsic chemical potential similar to passive systems, which depends on  density {\em and} self-propulsion speed, but {\em not} on the external potential, (ii) the external potential, and (iii) a newly derived one-body swim potential due to the activity of the particles. Our simulations on active Brownian particles show good agreement with our Fokker-Planck calculations, and confirm that $\mu(z)$ is spatially constant for several inhomogeneous active fluids in their steady states in a planar geometry. Finally,  we show that phase coexistence of ABPs with a planar interface satisfies not only mechanical but also diffusive equilibrium. The coexistence can be well-described by equating the bulk chemical potential and bulk pressure  obtained from bulk simulations for systems with low activity but requires explicit evaluation of the interfacial contributions at high activity.
\end{abstract}


\maketitle

\section{Introduction}
The non-equilibrium phase behavior of  active Brownian particles (ABPs), which constantly convert energy into directed motion, has received considerable attention in recent years.
The development of a thermodynamic framework to describe the clustering phenomena, the pronounced accumulation of active particles at walls, and the observed coexistence of dilute and dense phases of active matter that resemble gas-liquid and gas-solid coexistence in passive systems has been of particular interest\cite{fily2012athermal,redner2013structure,wysocki2014cooperative,bialke2012crystallization,takatori2014swim,farage2015effective,prymidis2015self,ni2013pushing,Solon2015,palacci2010sedimentation,Marconi2015,takatori2015towards,Stenhammar2014,enculescu2011active,Cates2013,Cates2015,vassilis}. Even the idea of basic thermodynamic variables such as temperature and pressure of these active systems is being heavily debated. For instance, the effective temperature introduced by Loi {\em et al.}\cite{Loi2011} and measured in experiments\cite{palacci2010sedimentation,enculescu2011active} was shown to depend not only on P\'eclet number, but also on the external potential and the particle interactions\cite{fily2012athermal,bialke2012crystallization,Cates2013,szamel2014self,takatori2015towards,ginot2015nonequilibrium,solon2015active,Takatori2016}.
Additionally, it was argued recently that the force per unit area on the wall can depend on the wall-particle interactions, which would imply that the pressure is not even a state function \cite{Solon2015,cates,Junot2017}.
Similarly, a chemical potential has been introduced in the literature using phenomenological arguments\cite{Stenhammar2014,Takatori2015,takatori2015towards,Solon2016}, or noise approximations \cite{Marconi2015} in an approach towards a thermodynamic framework for active systems. 
For instance, Takatori and Brady\cite{takatori2015towards} introduced a non-equilibrium chemical potential using micromechanical arguments, of similar form to the one that we will derive using the Fokker-Planck approach in this work. The authors of Ref.\cite{takatori2015towards}  even proceed and calculate spinodals and binodals on the basis of either a Gibbs-Duhem-like equation or a free energy for the (realistic) case of an incompressible solvent. Later, however, it was argued in Ref.\cite{Solon2015} that a Maxwell construction on the simulated equation of state does {\em not} yield the simulated coexistence densities. Consequently, a complete and well-established thermodynamic framework to describe the phase behavior of a model as simple as ABPs is still lacking. Our Fokker-Planck approach is similar in spirit to that of Ref.\cite{Solon2015,Solon2016}, but defines an expression for the local chemical potential in terms of the new concept of a ``swim potential'', which is well-defined in planar geometries and curl-free particle fluxes and which may contribute, in these cases, to formulating a theoretical framework.

In this study, we derive a microscopic expression for the local  chemical potential $\mu(z)$ of active Brownian particles in a spatially inhomogeneous steady state in a planar geometry, for simplicity, with $z$ the normal Cartesian direction.  We confirm using Brownian Dynamics simulations that $\mu(z)$ is  spatially constant for active fluids in contact with a soft planar wall, in a gravitational field, and in two-phase coexistence with a planar interface. Next, we show that the coexistence is described by diffusive and mechanical equilibrium with equal bulk pressure and bulk chemical potential of the coexisting phases, provided  the swim potential that we introduce in this article, is properly taken into account. However, we conclude that the swim potential and hence the chemical potential $\mu(z)$ is not a state function of the density for a macroscopic system.

\section{Methods and Formulation}
We consider a three-dimensional dispersion of $N$  active Brownian particles (ABPs) with positions $\mathbf{r}_i=(x_i,y_i,z_i)$ and orientations $\hat{\mathbf e}_i=(\sin \theta_i \cos \phi_i, \sin \theta_i \sin \phi_i, \cos \theta_i)$ with polar angle $\theta_i$ and azimuthal angle $\phi_i$, interacting via an isotropic pair potential  $V(|\mathbf{r}_i-\mathbf{r}_j|)$ and  subject to an external field $V_e(\mathbf{r}_i)$ for $i=1,\dots, N$ at temperature $T$. Particle $i$ experiences a constant self-propulsion force along its orientation $\hat{\mathbf e}_i$. 
The  motion of particle $i$   is described by the overdamped Langevin equations 
\begin{eqnarray}
\dot{\mathbf{r}}_i&=& -\beta D_t \nabla_i \left \lbrack V_e(\mathbf{r}_i)+\sum_{j\neq i} V(|\mathbf{r}_i  -\mathbf{r}_j|) \right \rbrack + v_0\hat{\mathbf e}_i +  \sqrt{2 D_{t}} {\mathbf \Xi}^t_i, \label{eos1} \\
\dot{\hat{\mathbf e}}_i &=& \sqrt{2 D_r} (\hat{\mathbf e}_i  \times \mathbf{\Xi}^r_i),  
\label{eos2}
\end{eqnarray}
where $D_{t}$ and $D_{r}$ are the translational and rotational diffusion coefficients,  $\beta=1/k_BT$ with $k_B$ the Boltzmann constant, and $v_0$ is the self-propulsion speed. The collisions with the solvent  are described by a stochastic force and torque characterised by random vectors  $\mathbf{\Xi}_i^t$ and $\mathbf{\Xi}_i^r$  with $\langle   \mathbf{\Xi}_i^t \rangle = \langle   \mathbf{\Xi}_i^r \rangle = \mathbf{0}$  and  $\langle \Xi^t_{i,\alpha}(t)\Xi^t_{j,\beta}(t') \rangle = \langle \Xi_{i,\alpha}^r(t)  \Xi_{j,\beta}^r(t') \rangle =  \delta_{\alpha\beta}\delta_{ij}\delta(t-t')$ with $\alpha,\beta=x,y,z$ \cite{fily2012athermal}. 

Starting from  (\ref{eos1}) and (\ref{eos2}), we  average over the noise  to derive the deterministic Fokker-Planck equation \cite{cates,Solon2015}
\begin{equation} 
\frac{\partial \psi(\mathbf{r},\hat{\mathbf e},t)}{\partial t} = -{\mathbf{\nabla}}\cdot  {\mathbf j}(\mathbf{r},\hat{\mathbf e},t) -  \mathbf{\nabla}_{\hat{\mathbf e}} \cdot {\mathbf j}_{\hat{\mathbf e}}(\mathbf{r},\hat{\mathbf e},t)
\label{FP}
\end{equation}
for the time evolution of the probability distribution function $\psi(\mathbf{r},\hat{\mathbf e},t) \equiv\langle \sum_{i=1}^{N} \delta(\mathbf{r}-\mathbf{r}_i)\delta({\hat{\mathbf e}}-{\hat{\mathbf e}_i}) \rangle$ with $\langle \dots \rangle$ the averaging over the random noise. Here we defined  the translational and rotational fluxes
\begin{eqnarray} 
{\mathbf j} =    &-&\beta D_t   \int \!\!\mathrm{d} {\mathbf r'} \int\!\!  \mathrm{d} \hat{\mathbf e}' \psi^{(2)}({\mathbf r},\hat{\mathbf e},\mathbf{r'},\hat{\mathbf e},t)  \mathbf{\nabla} V(|\mathbf{r} \! -\mathbf{r'}|) \nonumber \\ 
\mbox{} & &+ \Big (- \beta D_t\mathbf{\nabla} V_e(\mathbf{r})  + v_0 \hat{\mathbf e} \Big ) \psi(\mathbf{r},\hat{\mathbf e},t)  -D_t\mathbf{\nabla}   \psi(\mathbf{r},\hat{\mathbf e},t); \\
{\mathbf j}_{\hat{\mathbf e}}   = &-& D_r  \mathbf{\nabla}_{\hat{\mathbf e}}   \psi(\mathbf{r},\hat{\mathbf e},t). 
\end{eqnarray}
We introduced here the instantaneous full two-body correlation function $ \psi^{(2)}({\mathbf r},\hat{\mathbf e},\mathbf{r'},\hat{\mathbf e}',t)$   $\equiv  \langle  \sum_{i=1}^{N}\sum_{j\neq i}^N \delta(\mathbf{r}-\mathbf{r}_i)\delta({\hat{\mathbf e}}-{\hat{\mathbf e}_i})\delta(\mathbf{r}'-\mathbf{r}_j)\delta({\hat{\mathbf e}}'-{\hat{\mathbf e}_j})   \rangle$,  and hence to obtain a closed set of equations one needs a 
BBGKY-like hierarchy of Fokker-Planck equations  for the $n$-body correlation functions or a mean-field approximation  such as $\psi^{(2)}({\mathbf r},\hat{\mathbf e},\mathbf{r'},\hat{\mathbf e}',t) \simeq   \psi({\mathbf r},\hat{\mathbf e},t)  \psi(\mathbf{r'},\hat{\mathbf e}',t) $. 

The zeroth moment $\rho(\mathbf{r},t)=\int \mathrm{d} {\hat{\mathbf e}}~\psi(\mathbf{r},\hat{\mathbf e},t)$  defines the local particle density, and its time evolution is described by the continuity equation obtained from the zeroth moment of Eq.~(\ref{FP}),
\begin{equation}
\frac{\partial \rho(\mathbf{r},t)}{\partial t} = -\mathbf{\nabla} \cdot {\mathbf J}(\mathbf{r},t),  
\label{FP1}
\end{equation}
 with the particle flux ${\mathbf J}(\mathbf{r},t)= \int \mathrm{d} {\hat{\mathbf e}}~\mathbf{j}(\mathbf{r},\hat{\mathbf e},t)$ given by 
\begin{eqnarray}
\fl {\mathbf J}(\mathbf{r},t)\!=\! -\beta D_t \!\! \int \!\!\! \rho^{(2)}({\mathbf r},\mathbf{r'},t)\mathbf{\nabla} V(|\mathbf{r} \!\! -\mathbf{r'}|)   \mathrm{d} {\mathbf r'} \!-\! \beta D_t \!~\!\rho(\mathbf{r},t) \mathbf{\nabla} V_e(\mathbf{r})  \!+\!  v_0 {\mathbf m}(\mathbf{r},t) \!-\! D_t \mathbf{\nabla} \!\rho(\mathbf{r},t).
\label{flux}
\end{eqnarray}
Here $\rho^{(2)}({\mathbf r}, {\mathbf r'}, t) = \int\mathrm{d} {\hat{\mathbf e}} \int\mathrm{d} {\hat{\mathbf e}'} \psi^{(2)}({\mathbf r},\hat{\mathbf e},\mathbf{r'},\hat{\mathbf e}',t) $ is the spatial two-body correlation function and the first moment ${\mathbf m}(\mathbf{r},t) =  \int \mathrm{d} {\hat{\mathbf e}} ~\psi({\mathbf r},\hat{\mathbf e},t)\hat{\mathbf e}$   is the local polarization. 

An equation for ${\mathbf m}(\mathbf{r},t)$ follows from the first moment of Eq.~(\ref{FP}) which yields
\begin{eqnarray}
\frac{\partial {\mathbf m}({\mathbf r},t)}{\partial t} &=&  -\mathbf{\nabla} \!\cdot\!  {\cal J}_m({\mathbf r}, t) -(d-1) D_r {\mathbf m}({\mathbf r}, t), 
\label{FP2}
\end{eqnarray}
with  $d=2,3$ the spatial dimension of interest, and with the two-rank  momentum flux tensor
\begin{eqnarray}
\fl {\cal J}_m = -\beta D_t \!\!  \int \!\! \mathrm{d} {\hat{\mathbf e}} ~{\hat{\mathbf e}}\!\int \!\!\mathrm{d} {\mathbf r'}\!\! \int\!\mathrm{d} {\hat{\mathbf e}'} \psi^{(2)}({\mathbf r},{\hat{\mathbf e}'},{\mathbf r'},\hat{\mathbf e}',t)  \mathbf{\nabla} V(|\mathbf{r} \! -\mathbf{r'}|)  \nonumber \\
 -  \mathbf{m}({\mathbf r},t)  \beta D_t\mathbf{\nabla} V_e({\mathbf r}) + v_0 \Big( \rho({\mathbf r},t) ~\frac{\mathbb{I}}{d}+{\cal S}({\mathbf r},t)\Big) - D_t \mathbf{\nabla} {\mathbf m}({\mathbf r}, t ),
\end{eqnarray}
where ${\cal S}=\int \mathrm{d} \hat{\mathbf e}~\psi(\mathbf{r},\hat{\mathbf e})(\hat{\mathbf e}\hat{\mathbf e} -\mathbb{I}/d)$ is the traceless alignment tensor. 

We now assume that the system is only  inhomogeneous in the $z$-direction, due to either an external potential $V_e(z)$ or due to  coexistence of two phases separated by an interface  parallel to the $xy$-plane. 
Without loss of generality, we consider a large, but finite system by setting $V_e(\pm \infty)=\infty$, such that $\rho(z \rightarrow \pm \infty)=0$. 
From Eq.~(\ref{flux}), we find that the particle flux in the $z$-direction is given by 
\begin{eqnarray}
J_z(z,t) =  -\beta D_t& \int\!\! \mathrm{d} {\mathbf r'}  \rho^{(2)}({\mathbf r},{\mathbf r'},t) \partial_z V(|\mathbf{r} \! -\mathbf{r'}|) \nonumber \\
&-  \beta D_t \rho(z,t) \partial_z V_e(z)  +  v_0 m_z(z,t) - D_t \partial_z \rho(z,t).
\label{continuity}
\end{eqnarray}
When divided by $\beta D_t$, we interpret Eq.~(\ref{continuity}) as a continuum force balance rather than at the microscopic level, which requires averaging over bins that contain enough colloids for the continuum picture to hold. In the following sections this is achieved by having bins that are very elongated in the direction(s) perpendicular to the $z$-direction.

The term $(\beta D_t)^{-1}v_0m_z(z,t)$ has previously been interpreted as a contribution to the divergence of the stress tensor, which has led to a debate on pressure being a state function or not in active systems~\cite{cates,brady,speck2016ideal}. Here, however, we take another point of view, and regard this term as an activity-induced body force

\begin{equation}
-\rho(z,t) \partial_z V_{\mathrm{swim}}(z,t) \equiv \frac{v_0}{\beta D_t} m_z(z,t),
\label{swimpoteq}
\end{equation}
that is  exerted on the active particles by the solvent  \cite{brady,jeroen}.
This allows us to define the so-called swim potential  
\begin{equation}   
V_{\mathrm{swim}}(z,t) = V_{\mathrm{swim}}(z_0,t) -\! \frac{v_0}{\beta D_t } \int_{z_0}^z \frac{m_z(z',t)}{\rho(z',t)} \mathrm{d} z', 
\label{vswim}
\end{equation}
where  $V_{\mathrm{swim}}(z_0,t)$ is a suitably chosen reference.

Clearly,  for a homogeneous and isotropic bulk phase, for which the polarization ${\mathbf m}=0$ in a steady state,  $V_{\mathrm{swim}}$ is a spatial constant. Interestingly, however,  the value of this constant is determined by surfaces and interfaces, where ${\mathbf m}$ can be non-zero, not unlike the Donnan potential in inhomogeneous electrolyte solutions~\cite{verwey,zwanikken}. This is a reflection of the fact that the activity-induced body force on the active particles only averages out in the bulk, but not near  interfaces. 

We now combine Eqs.~(\ref{continuity})-(\ref{vswim}) to construct, in the spirit of the simplest dynamic density functional theory \cite{marconi1999dynamic,archer2009dynamical} with a density-independent diffusion coefficient, a local chemical potential-like function $\mu(z,t)$ by $J_z(z,t) = - D_t \rho(z,t) \partial_z \beta \mu(z,t)$ such that   
\begin{eqnarray}
\mu(z,t) - \mu(z_0,t) =& \mu_{\mathrm{int}}(z,t) - \mu_{\mathrm{int}}(z_0,t) \nonumber \\ 
&+ V_e (z)-V_e (z_0)
 + V_{\mathrm{swim}}(z,t)-V_{\mathrm{swim}}(z_0,t).
 \label{chempot1}
\end{eqnarray} 
The  external potential $V_e(z)$ and the intrinsic chemical potential $ \mu_{\mathrm{int}}(z,t)=k_BT\ln \rho(z,t)+ \mu_{\mathrm{ex}}(z,t)$, consisting of an ideal part and an excess chemical potential $\mu_{\mathrm{ex}}(z,t)$, are contributions  similar to those of a passive system. Here $\mu_{\mathrm{ex}}(z,t)$ is defined by
\begin{eqnarray}
\!\!\!\!\!\mu_{\mathrm{ex}}(z,t) =\mu_{\mathrm{ex}}(z_0,t) + \!\!\int_{z_0}^{z}\!\!\!\!\mathrm{d}z''\!\!\int \!\!\mathrm{d} {\mathbf r'} \rho(z',t)g(z'',z',R'',t) \partial_{z''} V(|\mathbf{r}'' \! -\mathbf{r'}|) , 
 \label{muex}
\end{eqnarray}
where we have used $ \rho^{(2)}({\mathbf r},{\mathbf r'},t)=\rho(z,t)\rho({z'},t)g(z,z',R,t)$, with the in-plane distance $R=\sqrt{(x-x')^{2}+(y-y')^2}$, in Eq.~(\ref{continuity}). Eq.~(\ref{chempot1}) reduces to the conventional chemical potential for a passive system, where $v_0=0$, and is constructed such that $J_z=0$ if $\mu(z,t)$ is a spatial constant. The local chemical potential $\mu(z)$ is therefore a prime candidate to describe diffusive equilibrium of coexisting phases in  stationary states of active systems. Interestingly, all terms in Eq.~(\ref{chempot1}) can be determined in Brownian Dynamics (BD) simulations of ABPs. 

The body-force interpretation of the polarization (\ref{swimpoteq}) can also be used to write the mechanical equilibrium condition of a stationary state in terms of  a well-defined  normal component  of the stress tensor. Since the stationary state satisfies $\partial \rho(z,t)/\partial t=0$, which from Eq.~(\ref{FP1}) is equivalent to  $J_z(z)=0$ for a macroscopically large, but finite system, we can rewrite Eq.~(\ref{continuity}) as 
 \begin{eqnarray}
\frac{\mathrm{d} P_N(z)}{\mathrm{d}z} + \rho(z) \partial_z V_{\mathrm{swim}}(z) = - \rho(z) \partial_z V_e(z)
\label{mech}
\end{eqnarray}
with the standard equilibrium-like expression for  the (intrinsic) normal pressure
\begin{eqnarray}
\fl P_N(z) =  P_{\mathrm{id}}(z) +  P_{\mathrm{vir}}(z) 
= \rho(z) k_BT - \int_{-\infty}^z \!\!\!\!\!\mathrm{d}z''\!\!\int^{\infty}_z \!\!\!\!\!\mathrm{d}z'\!\!\! \int \!\! \mathrm{d}\mathbf{R}' \!\!~\rho^{(2)}({\mathbf r}'',{\mathbf r'})  \partial_{z''} V(|\mathbf{r}'' \! -\mathbf{r'}|), 
 \label{normalpressure}
\end{eqnarray}
where we used Newton's third law and the symmetry of $\rho^{(2)}({\mathbf r},{\mathbf r'})$ under particle exchange. The last term in Eq.~(\ref{normalpressure}) is the virial contribution that describes the $z$-component of the interparticle forces across a plane at $z$, which can be measured in a BD simulation \cite{ikeshoji}. 
Note that  we did {\em not}   add a swim pressure \cite{brady,cates} to the ``intrinsic'' $P_N$, but instead treated  the activity at the level of a swim potential $V_{\mathrm{swim}}$ in the force balance (\ref{mech}), which turns out to be crucial for interpreting the (osmotic) pressure as a state function \cite{jeroen}. However, in order to connect to existing literature, and for later reference, we do define 
\begin{eqnarray}
\fl P_{\mathrm{swim}}(z)\!-\!P_{\mathrm{swim}}(z_0) \!=
\!\!\int_{z_0}^z   \!\!\!\rho(z') \partial_{z'} V_{\mathrm{swim}}(z') \mathrm{d}z'\!
=\frac{v_0 k_BT}{(d-1)D_tD_r} \left ({\cal J}_{m,zz}(z)-{\cal J}_{m,zz}(z_0)\right  )
\label{pswim}
\end{eqnarray}  
with the $zz$-component of ${\cal J}_{m}$ given by  
\begin{eqnarray}
\fl {\cal J}_{m,zz}(z) = \frac{v_0}{d} \rho(z) - m_z(z) \beta D_t \partial_z   V_e(z)  
 + v_0 {\cal S}_{zz}(z) -D_t \partial_z m_z(z)
\nonumber \\
-\beta D_t \int\!\! \mathrm{d} {\hat{\mathbf e}} \!\!\int \!\!\mathrm{d} {\mathbf r'} \!\!\int\!\!\mathrm{d} {\hat{\mathbf e}'} \psi^{(2)}({\mathbf r},\hat{\mathbf e},{\mathbf r'},\hat{\mathbf e}',t)  \partial_z V(|\mathbf{r} \! -\mathbf{r'}|) \cos \theta,
 \end{eqnarray}
which  reduces to the conventional swim pressure $P_{\mathrm{swim}}(z_b)= \rho(z_b)v_0^2 k_BT/(d(d-1)D_tD_r)$ in an ideal active bulk fluid at $z=z_b$ \cite{takatori2014swim,takatori2015towards}. Note that our local swim pressure   (\ref{pswim}) deviates from previous expressions  \cite{yang2014aggregation,winkler} due to the gradient term $\partial_z m_z$, which plays a non-negligible role in the force balance obtained from Eq.~(\ref{mech}) when significant spatial variations are present, e.g. in the interface of a phase coexistence. To summarize, we have introduced the concept of a swim potential here using a force balance for only the colloids. This force balance can be combined with an additional force balance for the solvent, which provides an alternative interpretation, but identical expression, for the swim pressure as an excess solvent pressure\cite{jeroen}.

With the definition (\ref{pswim}) one can thus define a total pressure $P(z) = P_N(z) + P_{\mathrm{swim}}(z)$, such that Eq.~(\ref{mech}) can be written as $\mathrm{d}P/\mathrm{d}z = - \rho(z) \partial_z V_e(z)$; in the case where $V_e(z)=0$ a steady state is then characterized by a spatially constant total pressure $P(z)$.
The intrinsic chemical potential $\mu_\mathrm{int}(z)$ and intrinsic normal pressure $P_N(z)$, and the swim potential $V_{\mathrm{swim}}(z)$ and swim pressure $P_{\mathrm{swim}}(z)$ have thus been constructed such that
\begin{equation}
  \frac{\mathrm{d}P_N(z)}{\mathrm{d}z} = \rho(z) \frac{\mathrm{d}\mu_{\mathrm{int}}(z)}{\mathrm{d}z},
  \quad \mathrm{and} \quad
  \frac{\mathrm{d}P_{\mathrm{swim}}(z)}{\mathrm{d}z} = \rho(z) \frac{\mathrm{d}V_{\mathrm{swim}}(z)}{\mathrm{d}z}.
  \label{gd1}
\end{equation}

If we now invoke a Local Density Approximation (LDA), i.e. assume that the local environment behaves as a bulk such that the local pressure and chemical potential are a function of only the local density $\rho(z)$, then Eq.~(\ref{gd1}) can be written in terms of bulk quantities as:
\begin{equation}
  \frac{\mathrm{d}P_N(\rho)}{\mathrm{d}\rho} = \rho \frac{\mathrm{d}\mu_{\mathrm{int}}(\rho)}{\mathrm{d}\rho},
  \quad \mathrm{and} \quad
  \frac{\mathrm{d}P_{\mathrm{swim}}(\rho)}{\mathrm{d}\rho} = \rho \frac{\mathrm{d}V_{\mathrm{swim}}(\rho)}{\mathrm{d}\rho},
\end{equation}
allowing us to write
\begin{eqnarray}
\frac{\mathrm{d} P(\rho)}{\mathrm{d}\rho} = \rho\frac{\mathrm{d} \mu(\rho)}{\mathrm{d}\rho}
\label{gd2}
\end{eqnarray}
with $\mu(\rho)=\mu_{\mathrm{int}}(\rho)+V_{\mathrm{swim}}(\rho)$, in a zero external potential. Here, we shall take care to distinguish the notation $\mu(\rho)$ for the chemical potential obtained via Eq.~(\ref{gd2}) from $\mu(z)$ which denotes the chemical potential calculated from Eq.~(\ref{chempot1}).
We recognize Eq.~(\ref{gd2}) as a generalization of the Gibbs-Duhem relation for equilibrium systems. Whereas in equilibrium (where $P_{\mathrm{swim}}=V_{\mathrm{swim}}=0$) it holds true in general, we emphasize that in this case we had to make use of the LDA to derive it. This Gibbs-Duhem relation provides a way to obtain the chemical potential $\mu(\rho)$ from the bulk equation of state $P(\rho)$, whereas to obtain $\mu(z)$ from Eq.~(\ref{chempot1}) we require complete spatial profiles. 
We test the applicability of Eq.~(\ref{gd2}) in simulations and show that it works well for cases with \emph{low anisotropy} (e.g. low polarization). However, Eq.~(\ref{gd2}) does not hold true in general as $V_{\mathrm{swim}}(z)\neq V_{\mathrm{swim}}^{\scriptscriptstyle LDA}(\rho(z))$ for \emph{high anisotropy} as we discuss later.

We note that Eq.~(\ref{gd2}) is akin to the one in Ref.\cite{takatori2015towards}, apart from a factor that is equal to the (incompressible) solvent volume fraction. The equilibrium analogue of Eq.~(\ref{gd2}) follows naturally if the solvent is treated grand-canonically  which we implicitly assume. Both approaches are also similar in the sense that they both identify the fluxes as being proportional to the gradient of a (scalar) chemical potential.

In the next Section, we apply the formalism of Eqs.~(\ref{vswim})-(\ref{pswim}) to active fluids and consider four different scenarios. We perform Brownian Dynamics (BD) simulations of non-interacting as well as interacting particles in two and three dimensions by employing Eqs.~(\ref{eos1}) and (\ref{eos2}). In Section~\ref{sec:results1} we study a non-interacting active fluid in contact with a short-ranged planar soft wall. We compare and verify that the stationary state is indeed described by  constant $\mu(z)$ in both the Fokker-Planck calculations and  particle based simulations. 
Next we present the results of BD simulations of an active fluid with Lennard-Jones (LJ) interactions subject to a gravitational field in Section~\ref{sec:results2}. In Section~\ref{sec:results3} we consider an active Lennard-Jones fluid exhibiting gas-liquid coexistence with a planar interface and confirm mechanical and diffusive equilibrium. We perform a Maxwell equal-area construction to identify phase coexistence from bulk equations of state. We then attempt to apply the same formalism to active particles which undergo Motility Induced Phase Separation at high activity in Section~\ref{sec:results4}.

\begin{figure}
	\flushright
	\includegraphics{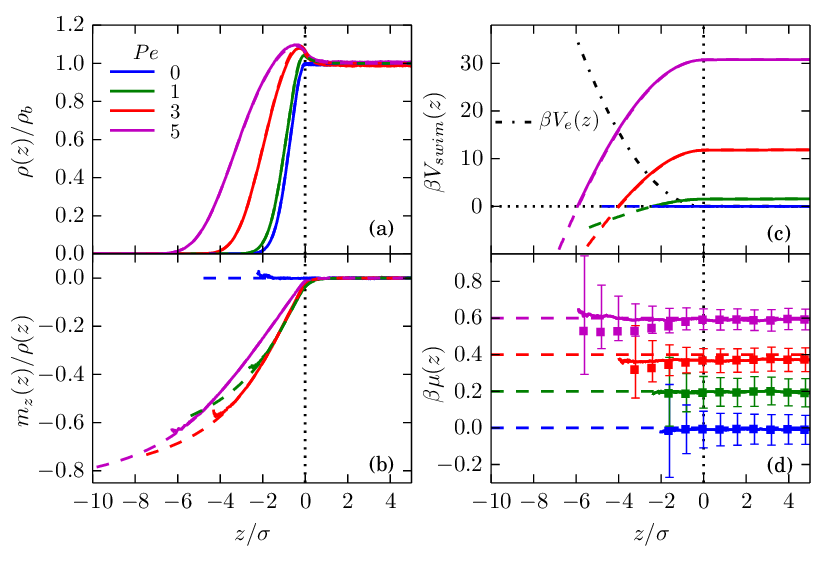}
	\caption{(a) Density profile $\rho(z)$, (b) polarization profile $m_z(z)/\rho(z)$, (c) swim potential $V_{\mathrm{swim}}(z)$ and the soft external potential $V_e(z)$ (see text), and (d) local chemical potential $\mu(z)$ (with error bars), all as a function of $z$ for an active ideal gas in contact with a planar soft wall, as obtained from BD simulations (solid lines) and Fokker-Planck calculations (dashed lines), for varying P\'eclet numbers as labeled. In (d) the solid lines represent $\beta\mu(z)$ obtained by the integration of $J_z(z)/\rho(z)$, which fluctuates about zero, whereas the square symbols show the resultant from Eq.~(\ref{chempot1}). The errorbars represent the error induced in $\beta\mu(z)$ due to the statistical error in $\rho(z)$. The deviation from the Fokker-Planck calculations deep into the wall for high Pe is due to the correlation of error upon integration.}
	\label{fig1}
\end{figure}

\section{Results}
\subsection{Active Ideal Gas}
\label{sec:results1}

We first consider a three-dimensional active ideal gas (with $V(r)=0$) at P\'eclet number Pe $=v_0/\sigma D_r=0$ (passive), 1, 3, 5, in the external potential $\beta V_e(z)= (z/\sigma)^2$ for $z<0$ and $V_e(z)=0$ for $z>0$, where the unit of length $\sigma=\sqrt{3D_t/D_r}$ is chosen to be the particle diameter so that the Stokes-Einstein relation for spheres in three dimensions is satisfied. Note that Pe can also be perceived as the ratio of the persistence length $v_0/D_r$ and the particle diameter\cite{takatori2014swim}.
For large  but finite $z=z_b \gtrsim 3\sigma$, the active fluid reaches a bulk state with bulk density $\rho_b=\rho(z_b)$, and the normal pressure  reduces to the bulk pressure $P_b=P_N(z_b)=  \rho_bk_BT$. 
In Fig.~\ref{fig1}(a) and (b) we show the time-averaged density profiles $\rho(z)$ and orientation profiles $m_z(z)/\rho(z)$, respectively. We observe that the particles penetrate deeper into the wall at higher Pe resulting into a more extended $\rho(z)$  within the wall accompanied by a small adsorption (that was found in Ref.\cite{Yan2015} as well) close to $z=0$. In Fig.~\ref{fig1}(b) we see no average polarization outside or inside the wall for the passive case. At finite Pe, however, Fig.~\ref{fig1}(b) shows that the average orientation is zero in the bulk where $V_e(z)=0$ and negative within the wall, corresponding to particles oriented towards the wall.
Fig.~\ref{fig1}(c) and (d) show $V_{\mathrm{swim}}(z)$ and $\mu(z)$ as obtained from Eq.~(\ref{vswim}) and (\ref{chempot1}), respectively. We find that $\mu(z)$ is indeed constant within our statistical accuracy of $\sim 0.1 k_BT$.
Clearly, for $\mu(z)$ to be constant  it is crucial that $V_{\mathrm{swim}}(z)$, which is attractive towards the wall consistent with the polarization and extended density profile close to the wall, is included in Eq.~(\ref{chempot1}); ignoring this contribution of $10-30 k_BT$ would {\em not} have yielded a spatially constant chemical potential in the stationary state. 
Although $\mu(z)$ was constructed to be spatially constant within the Fokker-Planck formalism, a confirmation from the simulations serves as a useful validation.

Additionally, we verify that the swim pressure (given by Eq.~(\ref{pswim})) measured in the bulk reduces to $P_{\mathrm{swim}}(z_b)= k_BTv_0^2\rho_b /(6D_rD_t)$. $V_{\mathrm{swim}}$ can similarly be obtained as $V_{\mathrm{swim}}(z_b)= (k_BT v_0^2/6 D_r D_t )\ln \rho_b\sigma^3$. We use this bulk state at $z_b\gtrsim 3 \sigma$ with $V_{\mathrm{swim}}(z_0=z_b)$ as the reference point for the profiles of $V_{\mathrm{swim}}(z)$ and $\mu(z)$ in Fig.~\ref{fig1}(c) and (d), respectively. 

\begin{figure}[h]
	\flushright
	\includegraphics{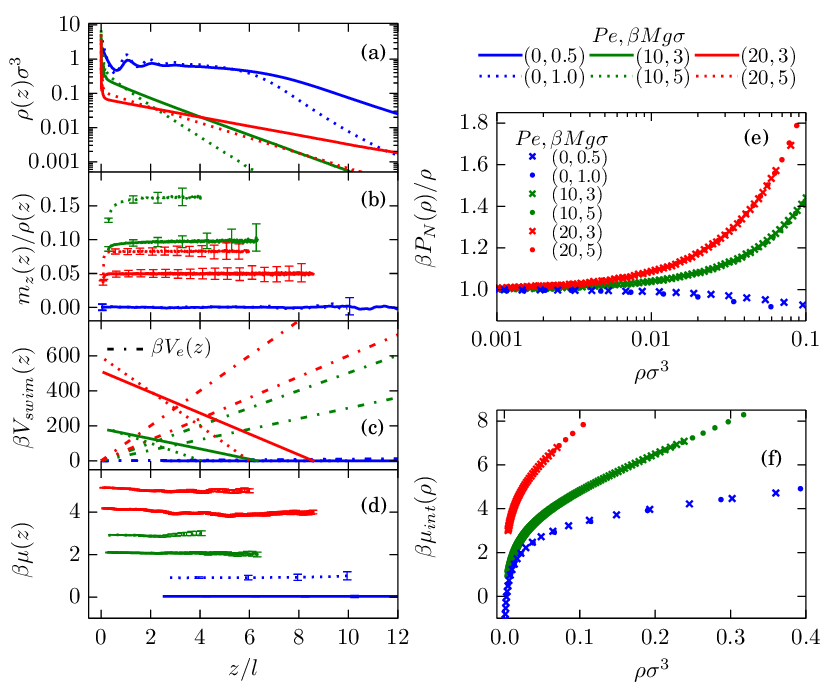}
	\caption{Height-dependence of (a) density $\rho(z)$, (b)~polarization $m_z(z)/\rho(z)$, (c)~swim potential $V_{\mathrm{swim}}(z)$, and (d)~chemical potential $\mu(z)$ (with an offset for clarity), all for an active LJ fluid in an external  gravitational potential $V_e(z) = Mgz$ for various values of $\beta Mg\sigma$, and P\'eclet number Pe=0 (blue), 10 (green), 20 (red) as obtained from BD simulations. The height $z$ is scaled with respect to $l$, where $l=v_0/D_r$ is the persistence length for Pe=10 and 20, and $l=\sigma$ is the particle diameter for Pe=0. The compressibility factor $P_N(\rho,{\textrm Pe})/\rho$ in (e) and the intrinsic chemical potential $\mu_{\mathrm{int}}(\rho,{\textrm Pe})$ shown with an offset in (f) show a proper collapse in the dilute limit for different $\beta Mg\sigma$ but {\em not} for Pe.}
	\label{fig2}
\end{figure}

\subsection{Sedimenting weakly active LJ-particles}
\label{sec:results2}

We now consider simulations of weakly active Lennard-Jones (LJ) particles with an isotropic pair potential, $V_{LJ}(r)=4\epsilon((\sigma/r)^{12}-(\sigma/r)^6)$, at $k_BT/\epsilon=2.0$ in the gravitational potential $V_e(z)=Mgz$ for $z>0$ with a hard `bottom' at $z=0$, with $M$ the buoyant particle mass. These systems are supercritical in the passive case, and therefore even more so in the active cases since the `critical temperature' decreases with increasing activity\cite{farage2015effective,vassilis}. 
We measure the density $\rho(z)\sigma^3$, polarization $m_z(z)/\rho(z)$, swim potential $\beta V_{\mathrm{swim}}(z)$, and chemical potential $\mu(z)$ for $\beta Mg\sigma=0.5$ and 1.0 for Pe=0, and $\beta Mg\sigma = 3$ and 5 for Pe=10 and 20, all plotted in Fig.~\ref{fig2}(a)-(d). 
In order to obtain a comparable length scale $l$ over which variations are observed in the passive (where we choose $l=\sigma$) and in the active cases (where $l=v_0/D_r$ ), we used a smaller buoyant mass of the particles in the passive case.
We observe that the polarization $m_z(z)$ is positive for Pe=10 and 20, and hence the  mean swimming direction is  opposite to the gravitational field, consistent with the findings in Ref.\cite{enculescu2011active}. 
Moreover, Fig.~\ref{fig2}(b) shows that the polarization profile $m_z(z)/\rho(z)$  is surprisingly constant over a large regime of heights $z$. 
As a consequence, the swim potential profile $\beta V_{\mathrm{swim}}(z)$ essentially decreases linearly with height $z$ for Pe $=10$ and 20 and counteracts largely the gravitational field, as shown in Fig.~\ref{fig2}(c), leading to an enormous increase in sedimentation length $(\beta Mg)^{-1}$\cite{palacci2010sedimentation}.
The chemical potential profile $\mu(z)$ is calibrated by $\mu(z_0)=0$ at the reference point $z_0$ determined by the condition $\rho(z_0)\sigma^3 = 5\times10^{-3}$. $\mu(z)$ is shown in Fig.~\ref{fig2}(d) and is indeed spatially constant within our statistical accuracy of $\sim 0.3 k_BT$. It is important to note here that $V_{\mathrm{swim}}(z)$ decreases by a few hundred $k_BT$ and the external gravitational potential $V_e(z) = Mgz$ increases by a few hundred $k_BT$ in the $z$-range of interest as shown in Fig.~\ref{fig2}(d).

In addition, we show in Fig.~\ref{fig2}(e) and (f) both $P_N$ and $\mu_{\mathrm{int}}$ as a function of $\rho$, obtained by eliminating $z$ from $P_N(z)$ and $\rho(z)$, and $\mu_{int}(z)$ and $\rho(z)$, respectively. We observe that the data collapse at fixed Pe, and it is alluring to interpret that $P_N(\rho,$Pe$)$ and $\mu_{\mathrm{int}}(\rho,$Pe$)$ are state functions of the density in this regime.

\begin{figure}
	\centering
	\includegraphics[height=0.6\textheight]{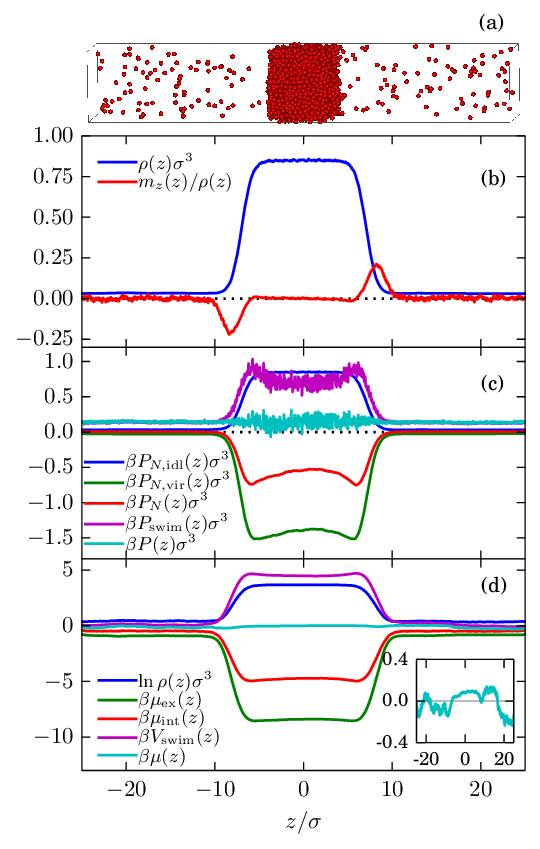}
	\caption{(a) A typical configuration of a three-dimensional gas-liquid coexistence of an active LJ fluid at Pe=3, and temperature $k_BT/\epsilon=0.43$, along with (b) the corresponding density profile $\rho(z)$ and polarization profile $m_z(z)/\rho(z)$, (c) total pressure $P(z)=P_N(z) + P_{\mathrm{swim}}(z)$ and the individual contributions, and (d) total chemical potential $\mu(z)$ obtained from Eq.~(\ref{chempot1}), and individual contributions, along with an inset showing a magnified view of $\mu(z)$. Both $P(z)$  and $\mu(z)$ are spatially constant within numerical accuracy,  demonstrating mechanical and diffusive equilibrium of the coexisting gas and liquid phase.}
	\label{fig3}
\end{figure}

\begin{figure}
	\flushright
	\includegraphics{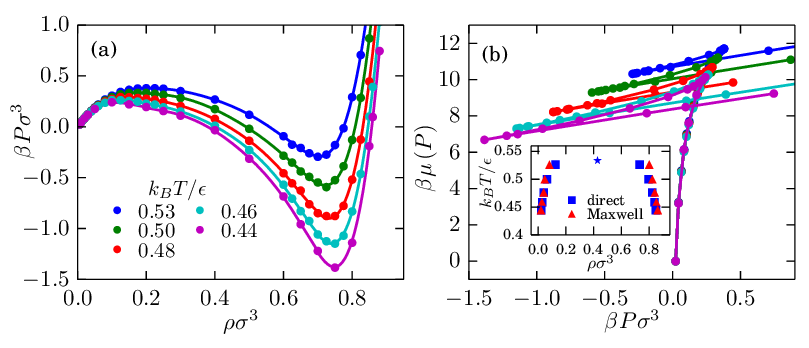}
	\caption{(a) Scaled pressure-density $P$-$\rho$, and (b) chemical potential-pressure  $\mu$-$P$ relations of an active LJ fluid at several temperatures $k_BT/\epsilon$ and Peclet Pe$=v_0/\sigma D_r=2.67$. The inset shows  the temperature-density gas-liquid binodals as obtained from direct coexistence simulations ({\color{blue} $\blacksquare$}) and from equating $\mu$ and $P$ in the coexisting phases ({\color{red} $\blacktriangle$}) of an active LJ fluid. }
	\label{fig4}
\end{figure}

\subsection{Active-LJ phase coexistence}  
\label{sec:results3}

We now consider a weakly active LJ fluid without any external potential ($V_e(z)=0$), and at subcritical temperatures such that coexistence of a gas and a liquid phase with bulk densities $\rho_g$ and $\rho_l$, respectively, is to be expected at overall intermediate densities $\rho_g < \rho <\rho_l$ in an elongated simulation box with periodic boundary conditions \cite{farage2015effective,vassilis}.
A temperature $k_BT/\epsilon=0.43$ and a P\'eclet number Pe$=v_0/D_r\sigma=3$ are used in this case.
In Fig.~\ref{fig3}(a), we show a typical configuration of a liquid slab in the center of the simulation box in coexistence with a gas phase on either side.
In Fig.~\ref{fig3}(b) we plot the corresponding density profile $\rho(z)$ which can be fitted to a hyperbolic tangent function (Eq.~(\ref{eqn:tanh_rho})), independently for $z>0$ and $z<0$, to obtain the coexistence densities $\rho(z_g)$ and $\rho(z_l)$ of the two bulk phases as fit parameters, with $z_g$ and $z_l$ a position in the bulk gas and liquid respectively. In the same figure we also plot the polarization profile $m_z(z)/\rho(z)$, showing that the swimming direction of the particles at the liquid-gas interface is pointing from the liquid phase towards the gas phase, i.e., against the attractive interparticle forces from the liquid\cite{vassilis,Paliwal2017}. 

In Fig.~\ref{fig3}(c) and (d) we plot the profiles $P(z)$ and $\mu(z)$, respectively, which clearly show that both are spatially constant.
We hence conclude that $P(z_g)=P(z_l)$ and  $\mu(z_g)=\mu(z_l)$, demonstrating mechanical and diffusive equilibrium of the coexisting gas and liquid phase. 
For completeness, in Fig.~\ref{fig3}(c) we also plot the individual contributions to the total pressure $P(z)=P_N(z) + P_{\mathrm{swim}}(z)$, where $P_{{\mathrm{swim}}}(z)$ is the swim pressure obtained from Eq.~(\ref{pswim}), and $P_N(z)=P_{N,\mathrm{idl}}(z)+P_{N,\mathrm{vir}}(z)$ is the normal pressure with the ideal pressure $P_{N,\mathrm{idl}}(z)$ and the virial contribution to the normal pressure $P_{N,\mathrm{vir}}(z)$ as obtained from Eq.~(\ref{normalpressure}). Similarly we plot the contributions to the chemical potential $\mu(z)=\mu_{\mathrm{int}}(z)+ V_{\mathrm{swim}}(z)$ in Fig.~\ref{fig3}(d), where the intrinsic chemical potential $\mu_{\mathrm{int}}(z)=k_BT \ln \rho(z) +\mu_{ex}(z) $ represents the sum of ideal and excess chemical potential. The swim potential $V_{\mathrm{swim}}(z)$ is calculated from the measured polarization profiles using Eq.~(\ref{vswim}).

In order to investigate if we can predict phase coexistence solely from bulk quantities, we perform BD simulations of bulk states of ABPs at several temperatures $k_BT/\epsilon$ and P\'eclet number Pe=2.67. We measure  the bulk pressure $P$ as a function of density $\rho$ in a simulation box small enough to prevent phase separation and plot the equations of state $P(\rho)$ for several subcritical temperatures in Fig.~\ref{fig4}(a). Now, within a Local Density Approximation (LDA), we apply the Gibbs-Duhem relation Eq.~(\ref{gd2}) and obtain $\mu(P)$ by integrating the equation of state $P(\rho)$ for several $T$'s  as shown in Fig.~\ref{fig4}(b). We emphasize here that we refer to $\mu(\rho)$ as the $\mu$ obtained by applying Eq.~(\ref{gd2}) which is not to be confused with $\mu(z)$. The intersection of the curve $\mu(P)$ gives the coexistence $\mu_g=\mu_l$ and $P_g=P_l$.
In the inset of Fig.~\ref{fig4}(b) we compare the binodals in the (scaled) temperature-density plane as obtained from the density profiles from direct coexistence simulations ($\rho(z_{g})$ and $\rho(z_{l})$) and from the bulk $\mu(P)$ intersections ($\rho_{g}$ and $\rho_{l}$). We find good agreement between the two results and thus conclude that the corresponding coexistence densities $\rho_g$ and $\rho_l$ could, in this (low Pe) case at least, be determined from the bulk equations of state.
Note that the activity has a huge effect on the gas-liquid binodals (shown in the inset of Fig.~\ref{fig4}(b)) as the critical temperature shifts from $k_BT/\epsilon\approx1.15$ in the passive case to $k_BT/\epsilon\approx0.54$ in the active case for Pe=2.67 (see Ref.~\cite{vassilis} for full comparison).

\begin{figure}
	\flushright
	\includegraphics{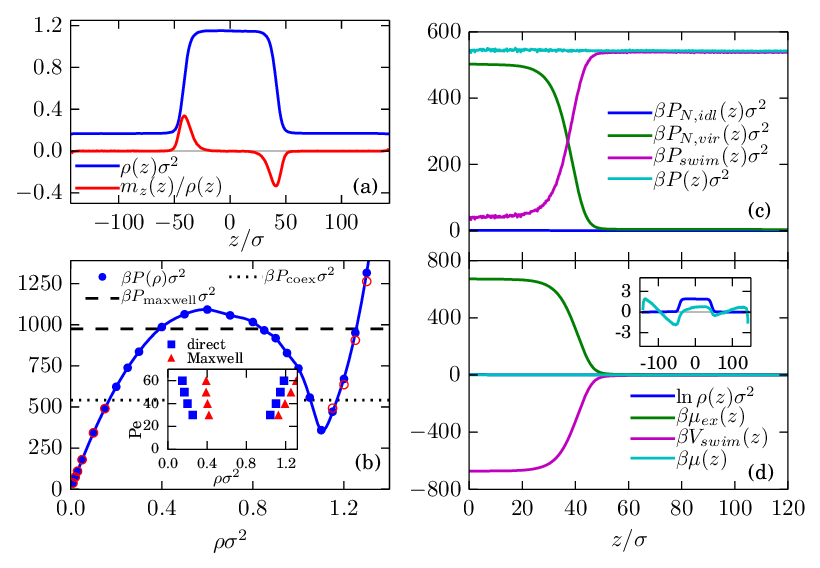}
	\caption{(a) Density $\rho(z)\sigma^2$ and polarization $m_z(z)/\rho(z)$ profiles of an active fluid with WCA interactions exhibiting MIPS at  $\mathrm{Pe}=50$, and temperature $k_BT/\epsilon=0.1$. (b) Pressure $\beta P(\rho)\sigma^2$ vs. density $\rho\sigma^2$ curve obtained from bulk simulations of small systems (solid circles) and large systems (open circles), together with Maxwell equal-area pressure (dashed line) and coexistence pressure $P_{\mathrm{coex}}=P(z)$ (dotted line) as measured in (c). The inset shows a comparison of bulk densities from direct coexistence simulations ({\color{blue} $\blacksquare$}) and the Maxwell equal-area construction ({\color{red} $\blacktriangle$}) for various Pe. (c)~Total pressure $P(z)=P_N(z) + P_{\mathrm{swim}}(z)$ profile, with the ideal, virial and swim contributions, and (d)~total chemical potential $\mu(z)$ profile with individual contributions, for $z>0$, corresponding to the system described in (a). The inset shows the ideal contribution $\beta \mu_{\mathrm{id}}(z)=\ln\rho(z)\sigma^2$ and that $\mu(z)$ is constant within an accuracy of $3k_BT$.}
	\label{fig5}
\end{figure}

\subsection{Motility Induced Phase Separation}
\label{sec:results4}
In this section we discuss the swim potential and the chemical potential in a two-dimensional system of strongly active particles exhibiting motility induced phase separation at high Pe. We choose our planar geometry in the $yz$ plane and assume homogeneity in the $y$ direction to be consistent with previous definitions. The particles interact with the WCA potential given by $V_{WCA}(r) = V_{LJ}(r) + \epsilon$, with a cut-off beyond $r\ge r_c = 2^{1/6}\sigma$ to make the particles purely repulsive. The particle orientations can be described in terms of a single angle $\theta_i$ as $\mathbf{\hat{e}}_i = (\cos\theta_i,\sin\theta_i)$.  The translational equation of motion in 2D is similar to Eq.~(\ref{eos1}) and the rotational diffusion follows $\dot{\theta}_i = \sqrt{2D_r}\Xi_i^r$, with $\Xi_i^r$ a zero-mean unit-variance Gaussian random variable.

As before, we fix rotational and translation diffusion coefficients to correspond to the particle interaction length scale $\sigma=\sqrt{3D_t/D_r}$ and change the self-propulsion speed $v_0$ to vary Pe. At high Pe, we find that the system phase separates into a gas phase and a dense phase, both of well-defined densities, separated by a planar interface in an elongated simulation box\cite{Cates2015}. For Pe=50 the typical density and polarization profiles are shown in Fig.~\ref{fig5}(a). Notably, the polarization profiles are now reversed with respect to Fig.~\ref{fig3}(b) as the particles at the interfaces are now pointing towards the dense phase. 
We measure the normal component of the total pressure $P(z)$ and the chemical potential $\mu(z)$ by summing the individual contributions, and plot them in Fig.~\ref{fig5}(c) and (d), respectively. We clearly observe that both the quantities $P(z)$ and $\mu(z)$ are spatially constant, demonstrating mechanical and diffusive equilibrium of the coexisting phases. With the polarization profiles reversed, $P_{\mathrm{swim}}(z)$ and $V_{\mathrm{swim}}(z)$ are now higher in the gas phase as compared to the denser phase.

Further, we perform a Maxwell equal-area construction on the equation of state. The $P-\rho$ curves shown in Fig.~\ref{fig5}(b) are obtained again using a small system size for which there is no global phase separation at intermediate densities. We confirm the results of the homogeneous states with larger system sizes and find that the agreement is satisfactory for our analysis. Performing a Maxwell construction on $P$ as a function of $1/\rho$ gives the equal-area pressure $P_{\mathrm{Maxwell}}$ shown as the dashed horizontal line in Fig.~\ref{fig5}(b). In the same figure, we also show the coexistence pressure $P_{\mathrm{coex}}$ obtained from the direct coexistence simulation of the phases coexisting at the corresponding set of parameters. From the two curves it is evident that the coexistence densities predicted by the Maxwell construction and the direct-coexistence simulations do \emph{not} agree. 
We perform the same procedure on a set of Pe in the range $30-60$ and plot the corresponding coexistence densities and the densities predicted by the Maxwell construction in the inset of Fig.~\ref{fig5}(b). From the disagreement between the two binodals we conclude that the Maxwell equal-area construction does not correspond to the coexisting states as obtained from the direct coexistence simulations, noted previously as well in Ref.\cite{Solon2015,Solon2016}. We have checked that using our $P(\rho)$ data with the definition of the chemical potential introduced in Ref.\cite{takatori2015towards} yields the same binodals as predicted here despite the difference of the factor concerning the solvent volume fraction.

\section{Discussion}
The results from the previous section show that the Maxwell equal-area construction, and hence the Gibbs-Duhem equation (20), cannot be used in general to predict the coexisting densities $\rho_g$ and $\rho_l$ \cite{Solon2015,Solon2016} in systems of ABPs.
In other words, even though $\mu(z_g)=\mu(z_l)$ in a phase-separated system (where $z_{g}$ and $z_{l}$ are locations far from interfaces such that the local densities are $\rho_{g}$ and $\rho_{l}$, respectively) , the chemical potentials obtained from the Gibbs-Duhem equation (\ref{gd2}) may not be equal, i.e. $\mu(\rho_g) \neq \mu(\rho_l)$.
The nonzero difference between $\mu(\rho_g)$ and $\mu(\rho_l)$ is caused by the failure of the LDA assumed in the derivation of Eq.~(\ref{gd2}), as we will show below. In particular, the values of $V_{\mathrm{swim}}(z)$ and $\mu_{ex}(z)$ in a bulk state at position $z$ and density $\rho_b$ do not only depend on $\rho_b$ (and other system parameters such as Pe) but \emph{also} on the interface between the bulk state and the reference state at $z_0$. This implies that neither $V_{\mathrm{swim}}$ nor $\mu_{ex}$ as expressed in Eqs.~(\ref{vswim}) and (\ref{muex}), respectively, are state functions of the density. Below we show an example for $V_{\mathrm{swim}}(z)$ which demonstrates this breakdown of the LDA in the case of a 2D active ideal gas (for which $\mu_{ex}(z)\equiv 0$) in a particular external potential.

\begin{figure}
	\centering
	\includegraphics{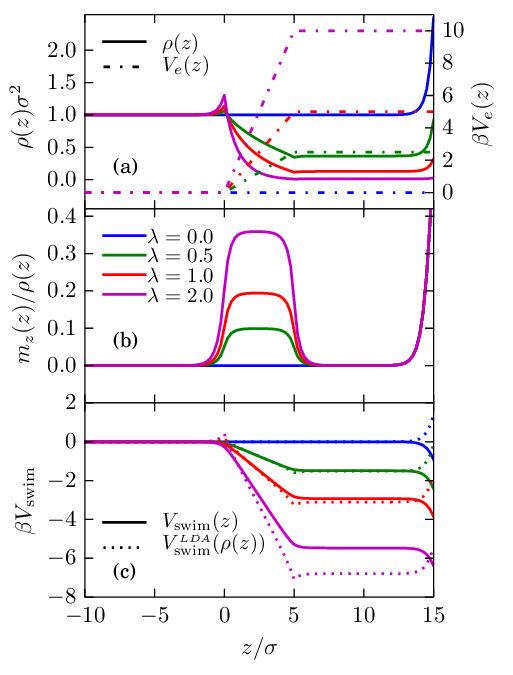}
	\caption{(a) Density profiles $\rho(z)$ and (b) polarization profiles $m_z(z)/\rho(z)$ of a non-interacting active fluid at Pe=1 in a ramp-shaped external potential with slope $\lambda=0,0.5,1,2$ shown as broken lines in (a). (c) Comparison of $V_{\mathrm{swim}}(z)$ obtained using Eq.~(\ref{vswim}) and $\beta V_{\mathrm{swim}}^{\scriptscriptstyle LDA}(\rho(z))=(v_0^2/2 D_tD_r)\ln\rho(z)\sigma^2$ obtained using LDA.}
	\label{fig6}
\end{figure}

The setup consists of a ramp-like external potential $\beta V_e(z)=\lambda z/\sigma$ in the region $0<z<5\sigma$ which separates a bulk region at the left (where $\beta V_e(z)=0$ for $z<0$) from the bulk on the right (where $\beta V_e(z)=5\lambda$ for $z>5\sigma$). These external potential are plotted in Fig.~\ref{fig6}(a) as dash-dot lines for $\lambda=0, 0.5, 1$, and 2.
The probability density $\psi(z,\theta)$ is obtained by solving Eq.~(\ref{FP}) for $V(r)=0$ numerically, at Pe $=1$ with a fixed density boundary condition $\rho\sigma^2=1.0$ for $z_0=-10\sigma$  and with a hard wall placed at $z=15\sigma$. The density and polarization profiles for increasing $\lambda$ are plotted in Fig.~\ref{fig6}(a) and (b), respectively.

\noindent In order to determine $V_{\mathrm{swim}}(z)$ for this non-interacting system with $V(r)\equiv 0$, Eq.~(\ref{pswim}) can be rewritten as
	\begin{eqnarray}
	\fl \!\!\!V_{\mathrm{swim}}(z)\!-\!V_{\mathrm{swim}}(z_0) \!
	&=\frac{v_0^2}{2\beta D_t D_r}\ln\left(\frac{\rho(z)}{\rho(z_0)}\right) \nonumber \\ &\quad +\!\frac{v_0}{\beta D_r}\!\!\int_{z_0}^{z}\!\! \frac{1}{\rho(z')} \frac{\mathrm{d}}{\mathrm{d}z'}\!\Bigg[\!\frac{v_0}{D_t} {\cal S}_{zz}(z')\!-\!\beta m_z(z')\frac{\partial V_e(z')}{\partial z'}\! -\! \frac{\partial m_z(z')}{\partial z'} \Bigg]\mathrm{d}z'.
	\label{eqn:vswim2}
	\end{eqnarray}
The $V_{\mathrm{swim}}(z)$ profiles, obtained equivalently from Eq.~(\ref{eqn:vswim2}) or from Eq.~(\ref{vswim}), are plotted as solid lines in Fig.~\ref{fig6}(c) where we have taken $z_0= -10\sigma$ as the reference state where $V_{\mathrm{swim}}(z_0)=0$. 
If we would approximate the vicinity of any point $z'$ as an isotropic bulk with density $\rho(z')$ in evaluating the swim potential $V_{\mathrm{swim}}(z')$, i.e. assume in Eq.~(\ref{eqn:vswim2}) ${\cal S}_{zz}(z') \approx m_z(z') \approx 0$ such that the term in square brackets vanishes for every $z'$, we obtain $\beta V_{\mathrm{swim}}^{\scriptscriptstyle LDA}(\rho(z))=(v_0^2/2D_t D_r)\ln\rho(z)\sigma^2$ which we refer to as the local density approximation (LDA) of Eq.~(\ref{eqn:vswim2}). Note that Eq.~(\ref{eqn:vswim2}) follows from the Fokker-Planck formalism, and this LDA does not refer to an approximation of a free-energy functional.
This $V_{\mathrm{swim}}^{\scriptscriptstyle {LDA}}(\rho(z))$, plotted as dotted lines in Fig.~\ref{fig6}(c), is equal to $V_{\mathrm{swim}}(\rho)$ obtained from the swim component of the Gibbs-Duhem-like relation (\ref{gd2}).
We find that $V_{\mathrm{swim}}(z)$ and $V_{\mathrm{swim}}^{\scriptscriptstyle LDA}(\rho(z))$ start to deviate at high $\lambda$ and do not coincide in the right bulk. Hence, we can conclude that the values for $V_{\mathrm{swim}}^{\scriptscriptstyle LDA}$ obtained from the Gibbs-Duhem equation are not correct in general.
This is due to the failure of LDA, i.e. due to the \emph{anisotropy} in the interface that renders the integral on the right hand side in Eq.~(\ref{eqn:vswim2}) non-negligible as compared to the first term. In Fig.~\ref{fig6}(b) we see that the polarization within the interface increases with $\lambda$, consistent with this idea of increasing anistropy. For an interacting system the forces between particles would add another contribution to $m_z(z)/\rho(z)$, which could also become a source of failure for the LDA. This will be studied in more detail in a future publication.

In Section~\ref{sec:results3} we observed that the Maxwell construction was able to predict the coexistence densities for the active LJ case with reasonable accuracy, but was in disagreement at higher activity in Section~\ref{sec:results4} for MIPS.
We now assert that the error made in the chemical potential by assuming the LDA translates into an error in the predicted coexisting densities that is small for the active LJ particles, but significant for MIPS.
We define the error in predicted coexistence densities of the gas and the dense phase, respectively, as $\Delta \rho^{err}_g=\rho(z_g)-\rho_g$ and $\Delta \rho^{err}_l=\rho(z_l)-\rho_l$, where $\rho(z_g)$ and $\rho(z_l)$ are the bulk coexistence densities and $\rho_g$ and $\rho_l$ denote the estimates obtained by performing a Maxwell construction. If we define the gas state as the reference state for the chemical potential, i.e. $\mu(z_0) = \mu(\rho_g)$ in Eq.~(\ref{chempot1}) with $z_0 = z_g$, then the error made in determining the chemical potential of the dense phase by using the Gibbs-Duhem equation (\ref{gd2}) is $\Delta \mu^{err}_l = \mu (\rho_l) - \mu(z_l)$, where we recall that $\mu(\rho_l)$ is the chemical potential of the dense phase obtained from the Gibbs-Duhem relation, whereas $\mu(z_l)$ is the true chemical potential determined in the coexistence simulation.  From $\Delta \mu^{err}_l$ the relative error in the predicted density of the dense phase can be estimated as $\Delta \rho^{err}_l / \rho(z_l) \approx 1 / \rho(z_l) \cdot \Delta \mu^{err}_l / (d\mu/d\rho)_l$. Similarly, the error in the predicted density of the gas phase can be estimated by using the dense phase as the reference state ($\mu(z_0)=\mu(\rho_l)$). The relative density error estimated in this manner is less than 5\% for the active LJ case, whereas it is of the order of 100\% for the MIPS case, which agrees with our findings in Fig.~\ref{fig4}(b) and \ref{fig5}(b), respectively.

We wish to make a note that the \emph{anisotropy} terms identified here resemble the interfacial contributions discussed in Ref.~\cite{Solon2016} for pairwise-interacting particles. Although it requires explicit measurement of these interfacial contributions by performing phase-coexistence simulations, Solon et al. were able to suggest a modified Maxwell construction for estimating the binodals in Ref.\cite{Solon2016}.

Moreover, our elongated simulation box in Section~\ref{sec:results3} and \ref{sec:results4} forces the system to  phase separate with a planar interface. Only for such a geometry $J_z(z)=0$, allowing us to write explicit expressions for mechanical and diffusive equilibrium. In other geometries the stationary state condition $\mathbf{\nabla} \cdot \mathbf{J}=0$ still allows for swirls that correspond to non-zero  $\mathbf{\nabla} \times \mathbf{J}$, for which our expressions for mechanical and diffusive equilibrium break down and a whole new framework is needed. Furthermore, the regime of applicability of  Eq.~(\ref{chempot1}) is limited by the underlying dynamic DFT  relation, where a $\rho$-independent diffusion coefficient $D_t$ is assumed; an extension to account for a $\rho$-dependent diffusion coefficient is left for a future study.

\section{Conclusions}
In conclusion, we have constructed expression (\ref{chempot1}) for the local chemical potential $\mu(z)$ for active fluids in a planar geometry, which includes the swim potential $V_{\mathrm{swim}}(z)$ defined by Eq.~(\ref{vswim}) in addition to ideal, excess, and external contributions well-known from equilibrium. Our BD simulations confirm that $\mu(z)$ is spatially constant in steady states of several inhomogeneous ideal and interacting fluids of active  particles, with  $V_{\mathrm{swim}}(z)$  an important contribution that counteracts either the external potential $V_e(z)$ or the excess contribution $\mu_{ex}(z)$. In the low activity regime studied for active LJ fluid, the chemical potential provides a method to predict the coexisting densities from bulk simulations. At high activity the \emph{anisotropy} in the interface causes the Gibbs-Duhem relation to be invalid, which provides support to the conclusions of Ref.\cite{Solon2016} that the details of the interface are necessary to determine the  coexisting bulk densities.
Our formalism opens new avenues towards a Fokker-Planck and dynamic density functional description (of stationary states) of active systems, especially for planar geometries.

\ack
S.P. and M.D. acknowledge funding from the Industrial Partnership Programme ``Computational Sciences for Energy Research'' (Grant No.14CSER020) of the Foundation for Fundamental Research on Matter (FOM), which is part of the Netherlands Organization for Scientific Research (NWO). This research programme is co-financed by Shell Global Solutions International B.V. This work is part of the D-ITP consortium, a program of the Netherlands Organisation for Scientific Research (NWO) that is funded by the Dutch Ministry of Education, Culture and Science (OCW). We acknowledge funding of an NWO-VICI grant. We would like to thank V. Prymidis for providing initial estimates and configurations for simulating MIPS in 2D, and also L. Filion and J. Tailleur for stimulating discussions.

\appendix
\section{Simulation details}
We perform Brownian Dynamics (BD) simulations for three-dimensional and two-dimensional geometries in Sections \ref{sec:results1}-\ref{sec:results3} and Section~\ref{sec:results4}, respectively. We use the Euler-Maruyama method to integrate the equations of motion (\ref{eos1}) and (\ref{eos2}) with a time step size $dt=10^{-5}\tau$ where $\tau=3Dr^{-1}$ is the unit of time. We keep the temperature of the bath fixed at $T$ by keeping the translational and rotational diffusion coefficients ($D_t$ and $D_r$ respectively) fixed and vary $v_0$ to change Pe and interaction strength $\epsilon$ to change the temperature of the colloidal particles.
We employ periodic boundary conditions in only  $x$- and $y$-direction in Section~\ref{sec:results1} and \ref{sec:results2}, in all three directions in Section~\ref{sec:results3}, and in both $y$- and $z$- directions in Section~\ref{sec:results4}. The system sizes are about 2500 particles in 3D and about 6500 particles in 2D for elongated box simulations. We measure the density profile $\rho(z)$ in the $z$-direction as $\rho(z)=\langle n(z)\rangle/L^2\Delta z$ by measuring the average of the number of particles $\langle n(z)\rangle$ in the slabs of volume $L^2\Delta z$ ($\rho(z)=\langle n(z)\rangle/L\Delta z$ in 2D) arranged parallel to $xy$ plane ($y$-direction in 2D), where $L$ is the length of the system in the $x$ and/or $y$-direction, and where $\Delta z = 0.1\sigma$ is the width of the slab. In a similar manner we measure the polarization profile $m_z(z)$ by summing the particle orientations in a slab at location $z$. The density profiles $\rho(z)$ can be fitted to a hyperbolic tangent function given by: 
	\begin{equation}   
	\rho(z) = \frac{1}{2} \left(\rho(z_l) + \rho(z_g)\right) +
				\frac{1}{2} \left(\rho(z_l) - \rho(z_g)\right) \tanh \left[\frac{2(z-z_0^*)}{D}\right], 
	\label{eqn:tanh_rho}
	\end{equation}
where $\rho(z_l)$ and $\rho(z_g)$ are the corresponding bulk liquid and vapour coexisting densities, $z_0^*$ is the location of the dividing plane and $D$ represents the thickness of the interface.
Subsequently, the swim potential profile $V_{\mathrm{swim}}(z)$ is obtained as
	\begin{equation}   
	V_{\mathrm{swim}}(z) = V_{\mathrm{swim}}(z_0) - \frac{v_0}{\beta D_t } \int_{z_0}^z \frac{m_z(z')}{\rho(z')} \mathrm{d} z', \label{vswimSI}
	\end{equation}
where we use $\rho(z)$ and $m_z(z)$ as measured in the BD simulations, and where $V_{\mathrm{swim}}(z_0)$ is a suitably chosen reference state. 
In addition, we measure the normal component of the stress tensor using 
	\begin{eqnarray}
	P_N(z) =  P_{\mathrm{id}}(z) +  P_{\mathrm{vir}}(z) 
	\end{eqnarray}
with the ideal gas pressure $P_{\mathrm{id}}(z)$ and the virial pressure $P_{\mathrm{vir}}(z)$ given by:
	\begin{eqnarray}
	 P_{\mathrm{id}}(z) =  \rho(z) k_BT \\
 	P_{\mathrm{vir}}(z) =   -  \frac{1}{2L^2\Delta z}\left \langle \sum_{i=1}^N\sum_{j \neq i}^N \frac{z_{ij}}{{r}_{ij}} \frac{\mathrm{d} V(r_{ij})}{\mathrm{d} r_{ij}} \int_{C_{ij}} {\hat z} \cdot \mathrm{d} {\mathbf l} \right \rangle,
 	\label{pvir}
	\end{eqnarray}
where ${r}_{ij} = |\mathbf{r}_{ij}| = |\mathbf{r}_{j}-\mathbf{r}_{i}|$ denotes the center-of-mass distance between particle $i$ and $j$, $z_{ij} = z_j -z_i$ where $z_i$ is the $z$ position of particle $i$, $C_{ij}$ is the intersection of $\mathbf{r}_{ij}$ and the slab of width $\Delta z$ centered at $z$. The integral in Eq.~(\ref{pvir}) denotes that the virial contribution to the pressure of particle pair $i$ and $j$  is due to the part of $\mathbf{r}_{ij}$ that lies inside the respective slab at $z$ within the coarse-grained Irving-Kirkwood approximation \cite{ikeshoji}. We also calculate the swim pressure 
	\begin{eqnarray}
	\fl P_{\mathrm{swim}}(z) =  \frac{v_0^2 k_BT}{d(d-1)D_tD_r} \rho(z) - \frac{v_0 m_z(z)}{(d-1)D_r}\partial_zV_e(z) \nonumber \\
	 -\frac{v_0 }{(d-1)D_r}\int \mathrm{d} {\mathbf {\hat e}} \int \mathrm{d} {\mathbf r'} \mathrm{d} {\mathbf {\hat e'}} \psi^{(2)}({\mathbf r},{\mathbf {\hat e}},{\mathbf r'},{\mathbf{\hat e'}})  (\nabla V(|\mathbf{r} -\mathbf{r'}|) \cdot \hat{z}) \cos \theta \nonumber \\
	 + \frac{v_0^2 k_BT}{(d-1)D_tD_r} {\cal S}_{zz}(z) - \frac{v_0 k_BT}{(d-1)D_r}\partial_z m_z(z), 
	\end{eqnarray}

\noindent and the chemical potential profile $\mu(z)$ using 
\begin{eqnarray}
\mu(z) = \mu(z_0) &+ k_BT \ln\rho(z) + \mu_{\mathrm{ex}}(z) + V_e (z) + V_{\mathrm{swim}}(z) \nonumber \\
	&-k_BT \ln\rho(z_0) - \mu_{\mathrm{ex}}(z_0)-V_e (z_0) -V_{\mathrm{swim}}(z_0),
\label{mu1}
\end{eqnarray} 
with the excess chemical potential $\mu_{\mathrm{ex}}(z)$ defined as 
\begin{eqnarray}
\mu_{\mathrm{ex}}(z) =\mu_{\mathrm{ex}}(z_0) + \!\!\int_{z_0}^{z} \!\!\mathrm{d}z' \left \langle \frac{1}{n(z')}\sum_{i=1}^{n(z')}\sum_{j \neq i}^N \frac{{z}_{ij}'}{{r}_{ij}'} \frac{\mathrm{d} V({r}_{ij}')}{\mathrm{d} {r}_{ij}'}  \right \rangle. 
\end{eqnarray}
Here, the excess chemical potential at $z$ with respect to a reference at $z_0$ is determined by integrating the averaged force that a particle feels due to the particle interactions with all other particles in the system over the distance $z_0$ to $z$.

\noindent Alternatively, if $V_e(z)=0$, $\mu(z)$ can also be obtained using 
\begin{eqnarray}
\mu(z) &=&\mu(z_0)+ \int_{z_0}^{z} \mathrm{d}z' \frac{1}{\rho(z')} \frac{\mathrm{d} P(z')}{\mathrm{d} z'} 
\label{mu2}
\end{eqnarray}
with $P(z) =  P_N(z)+ P_{\mathrm{swim}}(z)$.

\section*{References}
\bibliographystyle{iopart-num}
\bibliography{activeref}
\end{document}